\documentclass[prb,a4paper,showpacs,twocolumn]{revtex4}

\usepackage{amssymb}
\usepackage{amsmath}
\usepackage{amsfonts}
\usepackage{graphicx}

\DeclareMathOperator{\Sp}{sp} \DeclareMathOperator{\Tr}{Tr}
\DeclareMathOperator{\tr}{tr} 
\DeclareMathOperator{\sgn}{sgn}

\begin{document}

\title{Spin-valley interplay in two-dimensional disordered electron liquid}

\author{I.S.\,Burmistrov and N.M. Chtchelkatchev}
\affiliation{L.D.\ Landau Institute for Theoretical Physics,
Russian Academy of Sciences, 117940 Moscow, Russia}
\affiliation{Department of Theoretical Physics, Moscow Institute
of Physics and Technology, 141700 Moscow, Russia}
\begin{abstract}
We report the detailed study of the influence of the spin and
valley splittings on such physical observables of the
two-dimensional disordered electron liquid as resistivity, spin
and valley susceptibilities. We explain qualitatively the
nonmonotonic dependence of the resistivity with temperature in the
presence of a parallel magnetic field. In the presence of either
the spin splitting or the valley splitting we predict
\emph{novel}, with two maximum points, temperature dependence of
the resistivity.
\end{abstract}
\date{\today}


\maketitle


\section{Introduction\label{Sec:Intro}}

Disordered two-dimensional (2D) electron systems have been in the
focus of experimental and theoretical research for several
decades.~\cite{AFS} Recently, the interest to 2D electron systems
has been renewed because of the experimental discovery of
metal-insulator transition (MIT) in a high mobility silicon
metal-oxide-semiconductor field-effect transistor
(Si-MOSFET).~\cite{Pudalov1,prb95} Although, during last decade
the behavior of resistivity similar to that of
Ref.~[\onlinecite{Pudalov1,prb95}] has been found experimentally
in a wide variety of two-dimensional electron
systems,~\cite{Review} the MIT in 2D calls still for the
theoretical explanation.

Very likely, the most promising framework is provided by the
microscopic theory, initially developed by Finkelstein, that
combines disorder and strong electron-electron
interaction.~\cite{Finkelstein} Punnoose and
Finkelstein~\cite{LargeN} have shown possibility for the MIT
existence in the special model of 2D electron system  with the
infinite number of the spin and valley degrees of freedom. The
current theoretical results~\cite{BPS,BBP} do not support the MIT
existence for electrons without the spin and valley degrees of
freedom. Therefore, it is natural to assume that the spin and
valley degrees of freedom play a crucial role for the MIT in the
2D disordered electron systems.

Usually, in the MIT vicinity, from the metallic side, i.e., for an
electron density higher than the critical one, and at low
temperatures $T \ll \tau_\textrm{tr}^{-1}$ the initial increase of
the resistivity ($\rho$) with lowering temperature is replaced by
the decrease of $\rho$ as $T$ becomes lower than some sample
specific temperature.~\cite{Review} Here, $\tau_\textrm{tr}$
denotes the elastic scattering time. This nonmonotonic behavior of
the resistivity has been predicted from the renormalization group
(RG) analysis of the interplay between disorder and
electron-electron interaction in the 2D disordered electron
systems.~\cite{Finkelstein,FP} As a weak magnetic field $B$ is
applied parallel to the 2D plane, decrease of the resistivity is
stopped at some temperature and $\rho$ increases
again.~\cite{Pudalov2} Further increase of $B$ leads to the
monotonic growth of the resistivity as temperature is lowered,
i.e., to an insulating behavior, in the whole $T$-range. These
experimental results suggest the significance of the electron spin
for the existence of the metallic phase in the 2D disordered
electron systems.

As is well known, in both Si-MOSFET~\cite{AFS} and n-AlAs quantum
well~\cite{Shayegan0} 2D electrons can populate two valleys.
Therefore, these systems offer the unique opportunity for an
experimental investigation of an interplay between the spin and
valley degrees of freedom. Recently, using a symmetry breaking
strain to tune the valley occupation of the 2D electron system in
the n-AlAs quantum well, as well as a parallel magnetic field to
adjust the spin polarization, the spin - valley interplay has been
experimentally studied.~\cite{Shayegan1,Shayegan2} However, the
electron concentrations in the experiment were at least three
times larger than the critical one.~\cite{Shayegan0} Therefore,
the spin - valley interplay has been studied in the region of a
good metal very far from the metal-insulator transition.

In the present paper we report the detailed theoretical results on
the $T$-behavior of the 2D electron system with two valleys in the
MIT vicinity.  In particular, we study the effect of a parallel
magnetic field and/or a valley splitting ($\Delta_v$) on the
transport, and the spin and valley susceptibilities. We find that
in the presence of either the magnetic field or the valley
splitting the metallic behavior of the resistivity survives down
to the zero temperature.~\cite{JETPL1}  For example, this result
implies that at $B=0$ the metallic $\rho(T)$ dependence can be
observed experimentally at temperatures $T \ll \Delta_v$.  Only if
both the magnetic field and the valley splitting are present, then
the metallic behavior of the resistivity crosses over to the
insulating one. Next, we predict  novel, {\it with two maximum
points}, $T$-behavior of the resistivity in the presence of the
magnetic field and/or the valley splitting. Finally, we find that
as $T$ vanishes the ratio of the valley susceptibility ($\chi_v$)
to the spin one ($\chi_s$) becomes sensitive to the ratio of the
valley splitting to the spin one.  At high temperatures the ratio
$\chi_v/\chi_s$ is temperature independent and can be chosen equal
unity. If the spin splitting is larger (smaller) than the valley
splitting, then at low temperatures the ratio $\chi_v/\chi_s <(>)
1$. If  the spin and valley splittings are equal each other, then
the ratio $\chi_v/\chi_s = 1$ as temperature vanishes.

The presence of the parallel magnetic field and the
symmetry-breaking strain introduces new energy scales
$\Delta_s=g_L\mu_B B$ and $\Delta_v$ in the problem. Here, $g_L$
and $\mu_B$ stand for the $g$-factor and the Bohr magneton,
respectively. Let us assume that the following conditions hold:
$\Delta_v\ll \Delta_s\ll 1/\tau_\textrm{tr}$. In addition, a
magnetic field $B_\perp \gtrsim T/(D e)$ is applied perpendicular
to the 2D electron system in order to suppress the Cooper channel.
Here, $e$ and $D$ denote the electron charge and diffusion
coefficient, respectively. Due to the symmetry breaking, the spin
and valley splittings set the cut-off for a pole in the diffusion
modes (``diffusons'') with opposite spin and valley isospin
projections. In the temperature range $\Delta_s \ll  T \ll
\tau_\textrm{tr}^{-1}$ this cut-off is irrelevant and the 2D
electron system behaves as if no symmetry breaking terms are
applied. The temperature behavior of the resistivity is governed
by one singlet and $15$ triplet diffusive modes.~\cite{FP} At low
temperatures $\Delta_v\ll T\ll \Delta_s$, eight diffusive modes
with opposite spin projections do not contribute. Then, the
$\rho(T)$ dependence is determined by the remaining one singlet
and seven triplet modes. As we shall demonstrate below the
behavior of the resistivity can be either metallic or insulating.
Surprisingly, we found that the seven triplet diffusive modes are
not equivalent. They have to split into two groups of six and one
modes for the spin susceptibility  be $T$-independent.  For
temperatures $T\ll\Delta_v$, next four diffusive modes with
opposite isospin projections become ineffective. In this case, the
temperature dependence of the resistivity is determined by one
siglet and three triplet diffusive modes. Although, the number of
the remaining diffusive modes corresponds formally to
single-valley electrons with spin, the $\rho(T)$ behavior is
insulating.

The paper is organized as follows. In Section~\ref{Sec:Formalism}
we introduce the nonlinear sigma model that describes the
disordered interacting electron system. Then, we consider the
short length scales at which the system has $SU(4)$ symmetry in
the combined spin and valley space (Sec.~\ref{Sec:Short}). The
behavior of the system at the intermediate and long length scales
is studied in Sec.~\ref{Sec:Intermed} and Sec.~\ref{Sec:Low},
respectively. We end the paper with discussions of our results and
with conclusions (Sec.~\ref{Sec:Disc}).

\section{Formalism\label{Sec:Formalism}}

\subsection{Microscopic Hamiltonian\label{Sec:Formalism:MH}}

To start out, we consider 2D interacting electrons with two
valleys in the presence of a quenched disorder and a parallel
magnetic field at low temperatures $T\ll \tau_\textrm{tr}^{-1}$.
We assume that the perpendicular magnetic field $B_\perp \gtrsim
T/(D e)$ is applied in order to suppress the Cooper channel. Using
one electron orbital functions, we write an electron annihilation
operator as
\begin{equation}
\psi_\sigma(\mathbf{R}) = \sum_{\tau=\pm}
\psi^{\sigma}_{\tau}(\mathbf{r}) \varphi(z) e^{i\tau z
Q/2},\label{WF}
\end{equation}
where $z$ denotes the coordinate perpendicular to the 2D plane,
$\mathbf{r}$ the in-plane coordinate vector,  and $\mathbf{R} =
\mathbf{r}+z \mathbf{e_z}$. The subscript $\tau$ enumerates two
valleys and $\psi^\sigma_\tau$ is the annihilation operator of an
electron with the spin and isospin projections equal $\sigma/2$
and $\tau/2$, respectively. Let us assume that the wave functions
$\varphi(z)\exp (\pm i Q z/2)$ are normalized and orthogonal with
negligible overlap $\int dz \, \varphi^2(z) \exp(iQz)$. The vector
$\mathbf{Q}=(0,0,Q)$ corresponds to the shortest distance between
the valley minima in the reciprocal space: $Q\sim
a^{-1}_\textrm{lat}$, with $a_\textrm{lat}$ being the lattice
constant.~\cite{Iordansky,AFS}

In the path-integral formulation 2D interacting
electrons in the presence of the random potential $V(\mathbf{r})$
are described by the following grand partition function
\begin{equation}
Z = \int \mathcal{D}[\bar{\psi},\psi] e^{{S}[\bar{\psi},\psi]},
\end{equation}
with the imaginary time action
\begin{equation}
{S} = \int_0^{1/T} \!\!dt \Bigl
[-\bar{\psi}^{\sigma}_{\tau}(\mathbf{r},t)
\partial_t {\psi}^{\sigma}_{\tau}(\mathbf{r},t) -
\mathcal{H}_0-\mathcal{H}_\textrm{dis}-\mathcal{H}_\textrm{int}\Bigr
].\label{Hstart}
\end{equation}
The one-particle Hamiltonian
\begin{gather}
\mathcal{H}_0 = \int d\mathbf{r}\,
\bar{\psi}^{\sigma}_{\tau}(\mathbf{r}) \Bigl [
-\frac{\nabla^2}{2m_e}-\mu + \frac{\Delta_s}{2}\sigma    +
\frac{\Delta_v}{2} \tau \Bigr ] {\psi}^{\sigma}_{\tau}(\mathbf{r})
\end{gather}
describes a 2D quasiparticle with mass $m_e$ in the presence of
the parallel magnetic field and the valley splitting. Here, $\mu$
denotes the chemical potential. Next,
\begin{equation}
\mathcal{H}_\textrm{dis} = \int
d\mathbf{r}\,\bar{\psi}^{\sigma}_{\tau_1}(\mathbf{r})
V_{\tau_1\tau_2}(\mathbf{r}) {\psi}^{\sigma}_{\tau_2}(\mathbf{r})
\end{equation}
involves matrix elements of the random potential:
\begin{equation}
V_{\tau_1\tau_2}(\mathbf{r}) = \int dz\, V(\mathbf{R})
\varphi^2(z) e^{i (\tau_2-\tau_1)Q z/2}. \label{Veqr}
\end{equation}
In general, the matrix elements $V_{\tau_1\tau_2}(\mathbf{r})$   induce  both the intravalley and
intervalley scattering.
We suppose that
 $V(\mathbf{R})$ has the Gaussian
distribution, and
\begin{equation}
\langle V(\mathbf{R})\rangle = 0,\quad \langle
V(\mathbf{R}_1)V(\mathbf{R}_2)\rangle =
W(|\mathbf{r}_1-\mathbf{r}_2|,|z_1-z_2|),
\end{equation}
where the function $W$ decays at a typical distance $d$. If $d$ is
larger than the effective width of the 2D electron system, i.e.,
$d\gg [\int dz\,\varphi^4(z)]^{-1}$, then one can neglect the
$z$-dependence of $V(\mathbf{R})$ under the integral sign in
Eq.~\eqref{Veqr}. In this case, the intravalley scattering
survives only:
\begin{equation}
\langle
V_{\tau_1\tau_2}(\mathbf{r}_1)V_{\tau_3\tau_4}(\mathbf{r}_2)\rangle
= W(|\mathbf{r}_1-\mathbf{r}_2|,0) \delta_{\tau_1\tau_2}
\delta_{\tau_3\tau_4}.\label{W1}
\end{equation}
In the opposite case, $d\ll [\int dz\,\varphi^4(z)]^{-1}$, one
finds~\cite{Kuntsevich}
\begin{gather}
\langle
V_{\tau_1\tau_2}(\mathbf{r}_1)V_{\tau_3\tau_4}(\mathbf{r}_2)\rangle
=  \Bigl [ \delta_{\tau_1\tau_2} \delta_{\tau_3\tau_4}
\tilde{W}(|\mathbf{r}_1-\mathbf{r}_2|,0) \label{W2}
\\ +
\delta_{\tau_1\tau_4}\delta_{\tau_2,\tau_3}\delta_{\tau_1,-\tau_2}
\tilde W(|\mathbf{r}_1-\mathbf{r}_2|,2Q\tau_2) \Bigr ]\int
dz\,\varphi^4(z),\notag
\end{gather}
where $\tilde W(r,Q) = \int dz\, W(r,z)\exp(i Q z)$. The other
correlation functions, e.g. with $\tau_1=-\tau_2$ and
$\tau_3=\tau_4$, vanish due to integration over $(z_1+z_2)/2$
coordinate. It is the last term in Eq.~\eqref{W2} that contributes
to the intervalley scattering rate $1/\tau_v$. Assuming $Q^{-1}\ll
d$, one can neglect the intervalley scattering rate in comparison
with the intravalley scattering rate $1/\tau_i \sim
\tilde{W}(r,0)$. At last, allowing for a low electron
concentration $n_e$ in 2D electron systems, we consider the case
when the following inequality holds, $n_e d^2 \ll 1$. Then, both
Eqs.~\eqref{W1} and \eqref{W2} read
\begin{gather}
\langle
V_{\tau_1\tau_2}(\mathbf{r}_1)V_{\tau_3\tau_4}(\mathbf{r}_2)\rangle
= \frac{1}{2\pi\nu\tau_i}
\delta_{\tau_1\tau_2}\delta_{\tau_3\tau_4}\delta(\mathbf{r}_1-\mathbf{r}_2),\\
\frac{1}{\tau_i}=2\pi \nu \int d^2\mathbf{r}dz_1 dz_2\,
W(r,|z_1-z_2|) \varphi^2(z_1)\varphi^2(z_2), \notag \\ Q^{-1}\ll
d, \,[\int\varphi^4(z) dz]^{-1} \ll n_e^{-1/2}.\label{Condition1}
\end{gather}
Here, $\nu$ is the thermodynamic density of states. Under
conditions~\eqref{Condition1}, the interaction part of the
Hamiltonian is invariant under global $SU(4)$ rotations of the
electron operator $\psi^\sigma_\tau$ in the combined spin-valley
space:
\begin{equation}
\mathcal{H}_\textrm{int} =\frac{e^2}{2\epsilon} \int
d\mathbf{r}_1d\mathbf{r}_2\,\frac{\bar{\psi}^{\sigma_1}_{\tau_1}(\mathbf{r}_1)
{\psi}^{\sigma_1}_{\tau_1}(\mathbf{r}_1)
\bar{\psi}^{\sigma_2}_{\tau_2}(\mathbf{r}_2)
  {\psi}^{\sigma_2}_{\tau_2}(\mathbf{r}_2)}{|\mathbf{r}_1-\mathbf{r_2}|}.
\end{equation}
A dielectric constant of a substrate is denoted as $\epsilon$. The
low energy part of $\mathcal{H}_\textrm{int}$ can be written
as~\cite{Finkelstein,Castellani,KirkpatricBelitz,AleinerZalaNarozhny}
\begin{gather}
\mathcal{H}_\textrm{int} =\frac{1}{2} \int
d\mathbf{r}_1d\mathbf{r}_2\, \Bigl [ \rho(\mathbf{r}_1)
\Gamma_s(\mathbf{r}_1-\mathbf{r}_2) \rho(\mathbf{r}_2) \\ +
m^{a}(\mathbf{r}_1) \Gamma_t(\mathbf{r}_1-\mathbf{r}_2)
m^a(\mathbf{r}_2)\Bigr ]
\end{gather}
where
\begin{gather}
\rho(\mathbf{r}) = \sum_{\sigma\tau}
\bar{\psi}^\sigma_\tau(\mathbf{r}) {\psi}^\sigma_\tau(\mathbf{r}),
\\ m^a(\mathbf{r}) = \sum_{\sigma_1\sigma_2;\tau_1\tau_2}
\bar{\psi}^{\sigma_1}_{\tau_1}(\mathbf{r})
(t^a)^{\sigma_1\sigma_2}_{\tau_1\tau_2}
{\psi}^{\sigma_2}_{\tau_2}(\mathbf{r}). \notag
\end{gather}
Here, $\Gamma_s(\mathbf{q}) = U(q)+F_0^\rho/(4\nu)$ involves the
long-range part of the Coulomb interaction $U(q)=2\pi e^2/(q
\epsilon)$ and $\Gamma_t(\mathbf{q})=F_0^\sigma/(4\nu)$.
Quantities $F_0^{\sigma}$ and $F_0^{\rho}$ are the standard Fermi
liquid interaction parameters in the singlet and triplet channels,
respectively. The matrices $t^a$ with $a=1,\dots 15$ are the
non-trivial generators of the $SU(4)$ group.

\subsection{Nonlinear sigma model \label{Sec:Formalism:NLSM}}

At low temperatures, $T\tau_\textrm{tr}\ll 1$, the effective
quantum theory of 2D disordered interacting electrons described by
the Hamiltonian~\eqref{Hstart} is given in terms of the non-linear
$\sigma$-model. This theory involves unitary matrix field
variables
$Q^{\alpha_1\alpha_2;\sigma_1\sigma_2}_{mn;\tau_1\tau_2}(\mathbf{r})$
which obey the nonlinear constraint $Q^2(\mathbf{r})=1$. The
integers $\alpha_j=1,2,\dots,N_r$ denote the replica indices. The
integers $m, n$ correspond to the discrete set of the Matsubara
frequencies $\omega_n=\pi T (2n+1)$. The integers $\sigma_j=\pm 1$
and $\tau_j=\pm 1$ are spin and valley indices, respectively. The
effective action is
\begin{equation}
\mathcal{S} = \mathcal{S}_\sigma +\mathcal{S}_F+
\mathcal{S}_{sb}+\mathcal{S}_{vb}+\mathcal{S}_0,
\label{Sstart}
\end{equation}
where $\mathcal{S}_\sigma$ represents the free electron
part~\cite{FreeElectrons}
\begin{equation}
\mathcal{S}_\sigma = -\frac{\sigma_{xx}}{32} \Tr (\nabla Q)^2.
\label{SstartSigma}
\end{equation}
Here, $\sigma_{xx}$ denotes the mean-field conductivity in units
$e^2/h$. The symbol $\Tr$ stands for the trace over replica, the
Matsubara frequencies, spin and valley indices as well as
integration over space coordinates.

The term~\cite{Finkelstein}
\begin{gather}
\mathcal{S}_F = 4\pi T z \Tr\eta (Q-\Lambda)+\pi T\Gamma\int
d^2\mathbf{r}\sum_{\alpha n}\tr I_n^\alpha Q\tr I_{-n}^\alpha Q
\notag \\ - \pi T\Gamma_2\int d^2\mathbf{r}\sum_{\alpha n}(\tr
I_n^\alpha Q)\otimes(\tr I_{-n}^\alpha Q). \label{SstartF}
\end{gather}
involves the electron-electron interaction amplitudes which
describe the scattering on small ($\Gamma$) and large ($\Gamma_2$)
angles and the quantity $z$ originally introduced by
Finkelstein~\cite{Finkelstein} which is responsible for the
specific heat renormalization.~\cite{CasDiCas} The interaction
amplitudes are related with the standard Fermi liquid parameters
as~\cite{Finkelstein,Castellani,KirkpatricBelitz} $\Gamma_2 = -z
F_0^\sigma/(1+F_0^\sigma)$, $4 \Gamma = \Gamma_2+z
F_0^\rho/(1+F_0^\rho)$, and $z= \pi \nu^\star/2$ where
$\nu^\star=m^\star/(2\pi)$ with $m^\star$ being the effective
mass. The case of the Coulomb interaction corresponds to the
so-called ``unitary'' limit,~\cite{Aronov-Altshuler}
$F_0^\rho\to\infty$.

The symbol $\tr$ involves the same operations as in $\Tr$ except
the integration over space coordinates, and $\tr A \otimes \tr B =
A^{\alpha\alpha;\sigma_1\sigma_2}_{nn;\tau_1\tau_2}
B^{\beta\beta;\sigma_2\sigma_1}_{mm;\tau_2\tau_1}$. The matrices
$\Lambda$, $\eta$ and $I_k^\gamma$ are given as
\begin{gather}
\Lambda^{\alpha\beta;\zeta_1\zeta_2}_{nm} =
\mathrm{sign}\,(\omega_n)
\delta_{nm}\delta^{\alpha\beta}\delta^{\zeta_1\zeta_2}, \notag\\
\eta^{\alpha\beta;\zeta_1\zeta_2}_{nm} = n
\delta_{nm}\delta^{\alpha\beta}\delta^{\zeta_1\zeta_2},
\label{matrices_def}\\
(I_k^\gamma)^{\alpha\beta;\zeta_1\zeta_2}_{nm} =
\delta_{n-m,k}\delta^{\alpha\gamma}\delta^{\beta\gamma}\delta^{\zeta_1\zeta_2}.
\notag
\end{gather}
In the absence of $\Delta_s$ and $\Delta_v$, the action
$\mathcal{S}_\sigma+\mathcal{S}_F$ is invariant under the global
rotations
$Q_{nm;\tau_1\tau_2}^{\alpha\beta;\sigma_1\sigma_2}(\mathbf{r})
\to u_{\sigma_1\sigma_3}^{\tau_1\tau_3}
Q_{nm;\tau_3\tau_4}^{\alpha\beta;\sigma_3\sigma_4}(\mathbf{r})
[u^{-1}]_{\sigma_4\sigma_2}^{\tau_4\tau_2}$ in the combined spin-valley space for $u\in SU(4)$. The presence of the parallel
magnetic field and the valley splitting generates the symmetry
breaking terms:
\begin{equation}
\mathcal{S}_{sb}=  i z_s \Delta_s \Tr \sigma_z Q,\qquad
\mathcal{S}_{vb}= i z_v \Delta_v \Tr \tau_z Q, \label{Sstartsb}
\end{equation}
where $\mathbf{\sigma}_z$ and $\mathbf{\tau}_z$ are Pauli matrices
in the spin and valley spaces, respectively.  The $Q$-independent
part of the action reads~\cite{Finkelstein,Unify}
\begin{equation}
\mathcal{S}_0= -2\pi T z \Tr\eta\Lambda  + \frac{N_r}{2T}\int
d^2\mathbf{r} \left ( \chi_s^{0} \Delta_s^2 + \chi_v^{0}
\Delta_v^2\right ).\label{Sstart0}
\end{equation}
with $\chi_{s,v}^0 = 2 z_{s,v}/\pi$ being a bare value of the
spin (valley) susceptibility.

\subsection{$\mathcal{F}$-algebra \label{Sec:Formalism:Finv}}

The action~\eqref{Sstart} involves the matrices which are formally
defined in the infinite Matsubara frequency space. In order to
operate with them we have to introduce a cut-off for the Matsubara
frequencies. Then, the set of rules which is called
$\mathcal{F}$-algebra can be established.~\cite{Unify} At the end
of all calculations one should tend the cut-off to infinity.

The global rotations of $Q$ with the matrix $\exp(i \hat\chi)$
where $\hat \chi = \sum_{\alpha,n} \chi^\alpha_n I^\alpha_n$ play
the important role.~\cite{Unify,KamenevAndreev} For example,
$\mathcal{F}$-algebra allows us to establish the following
relations
\begin{eqnarray}
\Sp I^\alpha_n e^{i\hat\chi} Q e^{-i\hat\chi} &=& \Sp I^\alpha_n Q
+ 2 i n \chi^\alpha_{-n},\notag\\
\tr \eta e^{i\hat\chi} Q e^{-i\hat\chi} &=& \tr \eta Q +
\sum_{\alpha n} i n
(\chi^\alpha_n)^{\sigma_1\sigma_2}_{\tau_1\tau_2} \Sp I^\alpha_n
Q^{\sigma_2\sigma_1}_{\tau_2\tau_1} \notag \\ &-& \sum_{\alpha n}
n^2
(\chi^\alpha_n)^{\sigma_1\sigma_2}_{\tau_1\tau_2}(\chi^\alpha_{-n})^{\sigma_2\sigma_1}_{\tau_2\tau_1},\label{Falg}
\end{eqnarray}
where $\Sp$ stands for the trace over replica and the Matsubara
frequencies.

\subsection{Physical observables \label{Sec:Formalism:PhysObservables}}

The most significant physical quantities in the theory containing
information on its low-energy dynamics are physical observables
$\sigma_{xx}^\prime$, $z^\prime$, and $z_{s,v}^\prime$ associated
with the mean-field parameters $\sigma_{xx}$, $z$, and $z_{s,v}$
of the action~\eqref{Sstart}. The observable $\sigma_{xx}^\prime$
is the DC conductivity as one can obtain from the linear response
to an electromagnetic field. The observable $z^\prime$ is related
with the specific heat.~\cite{CasDiCas} The observables
$z_{s}^\prime$ and $z_{v}^\prime$ determine the static spin
($\chi_{s}^\prime$) and valley ($\chi_{v}^\prime$)
susceptibilities of the 2D electron
system~\cite{CastelChi,Finkelstein} as $\chi_{s,v}^\prime = 2
z_{s,v}^\prime/\pi$. Extremely important to remind that the
observable parameters $\sigma^\prime_{xx}$, $z_{s,v}^\prime$ and
$z^\prime$ are precisely the same as those determined by the
background field procedure.~\cite{RecentProgress}

The conductivity $\sigma^\prime_{xx}$ is obtained from
\begin{gather}
\sigma^\prime_{xx}(i\omega_n) = -\frac{\sigma_{xx}}{16 n}\left
\langle\tr[I_{n}^\alpha, Q][I_{-n}^\alpha, Q] \right
\rangle\hspace{1.5cm} \label{SigmaODef}
\\
+ \frac{\sigma_{xx}^2}{64 \mathbb{D} n } \int
d\mathbf{r}^\prime\langle\langle \tr I_n^\alpha
Q(\mathbf{r})\nabla Q(\mathbf{r}) \tr I_{-n}^\alpha
Q(\mathbf{r}^\prime)\nabla Q(\mathbf{r}^\prime)\rangle \rangle
\notag
\end{gather}
after the analytic continuation to the real frequencies, $i\omega_n \to
\omega + i0^+$ at $\omega \to 0$. Here, $\mathbb{D}=2$ stands for
the space dimension, and the expectations are defined with respect
to the theory~\eqref{Sstart}.

A natural definition of $z^\prime$
is obtained~\cite{Unify} through the derivative of the
thermodynamic potential $\Omega$ per the unit volume with respect
to $T$,
\begin{equation}
z^\prime = \frac{1}{2\pi \tr \eta \Lambda}\frac{\partial}{\partial
T}\frac{\Omega}{T}.\label{Defz}
\end{equation}
The observables $z_{s,v}^\prime$ are given as
\begin{equation}
z^\prime_{s,v} = \frac{\pi}{2 N_r} \frac{\partial^2
\Omega}{\partial \Delta_{s,v}^2} . \label{Defzsv}
\end{equation}

\section{$SU(4)$ symmetric case
\label{Sec:Short}}

\subsection{$\mathcal{F}$-invariance\label{Sec:Short:Finv}}

At short length scales $L\ll L_s,\,L_v$ where $L_{s,v} =
\sqrt{\sigma_{xx}/(16 z_{s,v}\Delta_{s,v})}$, the symmetry
breaking terms $\mathcal{S}_{sb}$ and $\mathcal{S}_{vb}$ can be
omitted and the effective theory becomes $SU(4)$ invariant in the
combined spin-valley space. Then, Eqs.~\eqref{SstartSigma} and
\eqref{SstartF} should be supplemented by the important constraint
that the combination $z+\Gamma_2-4\Gamma$ remains constant in the
course of the RG flow. Physically, it corresponds to the
conservation of the particle number in the
system.~\cite{Finkelstein} In the special case of the Coulomb or
other long-ranged interactions which are of the main interest for
us in the paper the relation
\begin{equation}
z+\Gamma_2-4 \Gamma=0\label{constraint}
\end{equation}
holds. With the help of Eqs.~\eqref{Falg}, one can check that
Eq.~\eqref{constraint} guarantees the so-called
$\mathcal{F}$-invariance~\cite{Unify} of the action
$\mathcal{S}_\sigma+\mathcal{S}_F$ under the global rotation of the
matrix $Q$:
\begin{equation}
Q(\mathbf{r}) \to e^{i\hat \chi} Q(\mathbf{r}) e^{-i\hat \chi},
\quad \hat \chi = \sum_{\alpha n} \chi^\alpha_n
I^\alpha_n.
\label{Finv}
\end{equation}
Here, $\chi^\alpha_n$ is the unit matrix in the spin-valley space. In
virtue of Eq.~\eqref{constraint}, it is convenient to introduce
the triplet interaction parameter $\gamma = \Gamma_2/z$ such that $\Gamma = (1+\gamma)
z/4$. We notice that the triplet interaction parameter is
related with $F_0^\sigma$ as $\gamma=-F_0^\sigma/(1+F_0^\sigma)$.

\subsection{Perturbative expansions \label{Sec:Short:Perturbative}}

To define the theory for the perturbative expansions we use the
``square-root'' parameterization
\begin{gather}
Q = W+\Lambda\sqrt{1-W^2},\qquad W =
  \begin{pmatrix}
    0 & w \\
    w^\dag & 0
  \end{pmatrix} .
\end{gather}
The action~\eqref{Sstart} can be written as the infinite series in
the independent fields
$w_{n_1n_2;\tau_1\tau_2}^{\alpha_1\alpha_2,\sigma_1\sigma_2}$ and
$w_{n_4n_3;\tau_1\tau_2}^{\dag
\alpha_1\alpha_2,\sigma_1\sigma_2}$. We use the convention that
the Matsubara frequency indices with odd subscripts $n_1, n_3, \dots$ run over
non-negative integers whereas those with even subscripts $n_2,
n_4, \dots$ run over negative integers. The propagators can be
written in the following form
\begin{widetext}
\begin{gather}
\langle
w_{n_1n_2;\tau_1\tau_2}^{\alpha_1\alpha_2;\sigma_1,\sigma_2}(\mathbf{p})
w_{n_4
n_3;\tau_4\tau_3}^{\dag\alpha_4\alpha_3;\sigma_4\sigma_3}(-\mathbf{p})
\rangle = \frac{16}{\sigma_{xx}}\,
\delta^{\alpha_1\alpha_3}\delta^{\alpha_2\alpha_4}
\delta_{n_{12},n_{34}} \Biggl \{
\delta^{\sigma_1\sigma_3}\delta^{\sigma_2\sigma_4}
\delta_{\tau_1\tau_3}\delta_{\tau_2\tau_4}\Bigl [ \delta_{n_1,n_3}
D_p(\omega_{12}) -\frac{32 \pi  T z\gamma}{\sigma_{xx}}
\delta^{\alpha_1\alpha_2}\notag
\\
\times D_p(\omega_{12})D^t_p(\omega_{12})\Bigr ]  + \frac{8 \pi T
z(1+\gamma)}{\sigma_{xx}} \delta^{\alpha_1\alpha_2}
\delta^{\sigma_1\sigma_3}\delta^{\sigma_2\sigma_4}
\delta_{\tau_1\tau_3}\delta_{\tau_2\tau_4}
D^s_p(\omega_{12})D^t_p(\omega_{12}) \Biggr \} ,\label{Prop}
\end{gather}
\end{widetext}
where $\omega_{12}=\omega_{n_1}-\omega_{n_2}$ and
\begin{gather}
D_p^{-1}(\omega_n) = p^2+\frac{16 z \omega_n}{\sigma_{xx}},\quad
[D^s_p(\omega_n)]^{-1} = p^2,
\\ [D^t_p(\omega_n)]^{-1} = p^2 +\frac{16
(z+\Gamma_2)\omega_n}{\sigma_{xx}}. \notag
\end{gather}

\subsection{Relation of $z_{s,v}$ with $z$ and $\gamma$\label{Sec:Short:Rel}}

The dynamical spin susceptibility $\chi_{s}(\omega,\mathbf{p})$
can be obtained from~\cite{Finkelstein}
\begin{equation}
\chi_s(i\omega_n,\mathbf{p}) = \chi_s^0 - T z_s^2 \langle \tr
I^\alpha_n \sigma_z Q(\mathbf{p}) \tr I^\alpha_{-n} \sigma_z
Q(-\mathbf{p})\rangle \label{SS1}
\end{equation}
by the analytic continuation to the real frequencies, $i\omega_n \to
\omega + i0^+$. Similar expression is valid for the valley
susceptibility. Evaluating Eq.~\eqref{SS1} in the tree level
approximation with the help of Eqs.~\eqref{Prop}, we obtain
\begin{equation}
\chi_s(i\omega_n,\mathbf{p}) = \frac{2 z_s}{\pi} \left (1 -
\frac{16 z_s \omega_n}{\sigma_{xx}} D^t_p(\omega_n)\right ).
\end{equation}
In the case
$\Delta_s=\Delta_v=0$ the total spin conserves, i.e., $\chi(\omega,\mathbf{p}=0)=0$.
In order to be consistent with this physical requirement, the relation
\begin{equation}
z_s = z+\Gamma_2\equiv z(1+\gamma)\label{zsR}
\end{equation}
should hold. Similarly, the total valley isospin conservation
guarantees that
\begin{equation}
z_v = z+\Gamma_2\equiv z(1+\gamma).\label{zvR}
\end{equation}
Being related with the conservation laws, Eqs.~\eqref{zsR} and
\eqref{zvR} are valid also for the observables:
\begin{equation}
z_{s}^\prime=z_v^\prime=z^\prime(1+\gamma^\prime).
\end{equation}
Therefore,
three physical observables $\sigma_{xx}^\prime$, $z^\prime$ and
$\gamma^\prime$ completely determines the renormalization of the
theory~\eqref{Sstart} at short length scales $L \ll L_s, L_v$.

\subsection{One loop renormalization group equations \label{Sec:Short:1loop}}

As is shown in Ref.~[\onlinecite{FP}], the standard one-loop analysis
for the action $\mathcal{S}_\sigma+\mathcal{S}_F$ yields the
following renormalization group functions that determine the
zero-temperature behavior of the observable parameters with
changing of the length scale $L$
\begin{eqnarray}
\frac{d\sigma_{xx}}{d \xi} &=& \beta_\sigma=-\frac{2}{\pi} \left
[1+15 f(\gamma) \right ] ,\label{RG0_1}\\ \frac{d\gamma}{d\xi} &=&
\beta_{\gamma}=\frac{(1+\gamma)^2}{\pi\sigma_{xx}},\label{RG0_2}\\
\frac{d\ln z}{d \xi} &=&
\gamma_z=\frac{15\gamma-1}{\pi\sigma_{xx}}. \label{RG0_3}
\end{eqnarray}
Here, $f(\gamma)=1-(1+\gamma^{-1})\ln(1+\gamma)$, $\xi = \ln L/l$
and we omit prime signs for a brevity. Physically, the microscopic
length $l$ is the mean-free path length. It is the length at which
the bare parameters of the action~\eqref{Sstart} are defined.
Renormalization group Eqs.~\eqref{RG0_1}-\eqref{RG0_3} are valid
at short length scales $L \ll L_s,\,L_v$.

As is well-known,~\cite{FP} solution of the RG
Eqs.~\eqref{RG0_1}-\eqref{RG0_2} yields the dependence of the
resistivity $\rho=1/\pi \sigma_{xx}$ on $\xi$ which has the
maximum point and $\gamma(\xi)$ dependence that monotonically
increases with $\xi$.

\section{$SU(2)\times SU(2)$ symmetry case
\label{Sec:Intermed}}

\subsection{Effective action \label{Sec:Intermed:Effect}}

In this and next sections we assume that the spin splitting is much larger than the valley splitting,
$\Delta_s\gg\Delta_v$. Then,
at intermediate length scales $L_s\ll L \ll L_v$ the symmetry
breaking term $\mathcal{S}_{sb}$ becomes important. In the quadratic
approximation it reads
\begin{equation}
\mathcal{S}_{sb} = \frac{i z_s\Delta_s}{2}\int d\mathbf{r}
\sum^{\alpha_j,\sigma_j}_{n_j,\tau_j}
 \left (\sigma_2-\sigma_1\right
 )w_{n_1n_2;\tau_1\tau_2}^{\alpha_1\alpha_2;\sigma_1\sigma_2}
 \bar{w}_{n_2n_1;\tau_2\tau_1}^{\alpha_2\alpha_1;\sigma_2\sigma_1}
\end{equation}
Hence, the modes in
$Q^{\alpha\beta;\sigma_1\sigma_2}_{nm;\tau_1\tau_2}$ with
$\sigma_1\neq \sigma_2$ acquire a finite mass of the order of
$z_s\Delta_s$ and, therefore, are negligible at length scales
$L\gg L_s$. As a result, $Q$ becomes a diagonal matrix in the spin
space. Then, the spin susceptibility has no renormalization on these
length scales, i.e,
\begin{equation}
\frac{d z_s}{d \xi} = 0,\qquad L_s\ll L\ll L_v.
\end{equation}

Let us denote $Q^{\alpha\beta;\pm 1\pm
1}_{nm;\tau_1\tau_2}=[Q^{\alpha\beta}_{nm;\tau_1\tau_2}]_\pm$.
Then, the action~\eqref{Sstart} becomes
$\mathcal{S}=\mathcal{S}_\sigma+\mathcal{S}_F+\mathcal{S}_{vb}$
where
\begin{equation}
\mathcal{S}_\sigma = -\frac{\sigma_{xx}}{32} \sum_{\sigma=\pm}\int
d^2\mathbf{r} \tr (\nabla Q_\sigma)^2 \label{SstartSigma1}
\end{equation}
and
\begin{eqnarray}
\mathcal{S}_F &=& 4\pi T z \sum_\sigma\int d^2\mathbf{r}\tr\eta
(Q_\sigma-\Lambda)\label{SstartF1}
\\&+&\pi T\int d^2\mathbf{r}\sum_{\alpha n}\sum_{\sigma_1,\sigma_2=\pm}
\Gamma_{\sigma_1\sigma_2} \tr I_n^\alpha Q_{\sigma_1}\tr
I_{-n}^\alpha Q_{\sigma_2} \notag
\\ &-& \pi T\Gamma_2\int d^2\mathbf{r}\sum_{\alpha
n}\sum_{\sigma=\pm}(\tr I_n^\alpha Q_\sigma)\otimes(\tr
I_{-n}^\alpha Q_\sigma).\notag
\end{eqnarray}
Now, the symbol $\tr$ stands for the trace over replica,
the Matsubara frequencies, and the valley indices whereas $\Tr = \int d^2 \mathbf{r}
\tr$. The action~\eqref{SstartF1} corresponds to the following low
energy part of the Hamiltonian describing electron-electron interactions:
\begin{gather}
\mathcal{H}_\textrm{int} = \frac{1}{2} \int d\mathbf{r} \Bigl [
\sum_{\sigma_1,\sigma_2} \rho^{\sigma_1}
\Gamma_s^{\sigma_1\sigma_2}\rho^{\sigma_2} +  m^a\Gamma_t m^a
\Bigr ], \label{HintSB}
\\
\rho^\sigma = \sum_\tau \bar{\psi}^\sigma_\tau \psi^\sigma_\tau,
\qquad m^a = \sum_{\sigma\tau\tau^\prime} \bar{\psi}^\sigma_\tau
(t^a)_{\tau\tau^\prime}^{\sigma\sigma}
\psi^\sigma_{\tau^\prime}\notag
\end{gather}
It is worthwhile mentioning that Eq.~\eqref{HintSB} is in
agreement with the ideas of Ref.~[\onlinecite{Meerovich,Zala}].

The
symmetry breaking part reads
\begin{equation}
\mathcal{S}_{sb}=  i z_v \Delta_v \sum_{\sigma=\pm} \int
d^2\mathbf{r} \tr \tau_z Q_\sigma. \label{Sstartsb1}
\end{equation}
At length scales $L \sim L_s$, the couplings
$\Gamma_{\sigma_1\sigma_2}$ are all equal to each other,
$\Gamma_{\sigma_1\sigma_2}(L\sim L_s)=\Gamma= (z+\Gamma_2)/4$.
However, the symmetry allows the following matrix structure of
$\hat\Gamma$:
\begin{gather}
\hat\Gamma = \begin{pmatrix}
  \Gamma_{++} & \Gamma_{+-} \\
  \Gamma_{+-} & \Gamma_{++}
\end{pmatrix}.
\end{gather}
As we shall see below, this matrix structure is consistent with
the renormalization group. Physically, $\Gamma_{++}$ and
$\Gamma_{+-}$ describe interactions between electrons with the
same and opposite spins, respectively.

The action~\eqref{SstartSigma1} and \eqref{SstartF1} is invariant
under the global rotations
$[Q_{nm;\tau_1\tau_2}^{\alpha\beta}]_\sigma(\mathbf{r}) \to
u_\sigma^{\tau_1\tau_3}
[Q_{nm;\tau_3\tau_4}^{\alpha\beta;}]_\sigma(\mathbf{r})
[u^{-1}]_\sigma^{\tau_4\tau_2}$ in the valley space for
$u_\sigma\in SU(2)$. In order to preserve the invariance under the
global rotations
\begin{equation}
Q_\pm(\mathbf{r}) \to e^{i\hat \chi} Q_\pm(\mathbf{r}) e^{-i\hat
\chi}, \quad \hat \chi = \sum_{\alpha n} \chi^\alpha_n
I^\alpha_n,\label{Finv1}
\end{equation}
where $\chi^\alpha_n$ is the unit matrix in the valley space, the
following relation has to be fulfilled
\begin{equation}
z + \Gamma_2 - 2 \Gamma_{++} = 2\Gamma_{+-}.\label{constraint1}
\end{equation}
Physically, this equation corresponds to the particle number conservation and
is completely analogous to Eq.~\eqref{constraint}.

\subsection{Perturbative expansions \label{Sec:Intermed:Pert}}

In order to resolve the constraint $Q_\pm^2=1$ we use the
``square-root'' parameterization:
\begin{equation}
Q_\pm = W_\pm+\Lambda\sqrt{1-W_\pm^2}.
\end{equation}
Then, the action~\eqref{SstartSigma1} and \eqref{SstartF1}
determines the propagators as follows
\begin{equation}
\langle [w_{n_1n_2}^{\alpha_1\alpha_2;\tau_1,\tau_2}(q)]_\sigma
[w_{n_4
n_3}^{\dag\alpha_4\alpha_3;\tau_4\tau_3}(-q)]_{\sigma^\prime}
\rangle = \frac{32}{\sigma_{xx}}
\hat{\mathcal{D}}_{\sigma\sigma^\prime},
\end{equation}
where
\begin{eqnarray}
& &\hat{\mathcal{D}} =
\delta^{\alpha_1\alpha_3}\delta^{\alpha_2\alpha_4}\delta_{n_{12},n_{34}}
\Bigl [\delta_{n_1,n_3}\delta^{\tau_1\tau_3}\delta^{\tau_2\tau_4}
D_q(\omega_{12},\tau_1,\tau_2) \notag \\&& -\frac{32\pi
T}{\sigma_{xx}}\Gamma_2
\delta^{\alpha_1\alpha_2}\delta^{\tau_1\tau_3}\delta^{\tau_2\tau_4}
D_q(\omega_{12},\tau_1,\tau_2) D^t_q(\omega_{12},\tau_1,\tau_2)
\notag
\\&&+ \frac{32\pi T}{\sigma_{xx}}\hat \Gamma
\delta^{\alpha_1\alpha_2}\delta^{\tau_1\tau_2}\delta^{\tau_3\tau_4}
\hat D^s_q(\omega_{12}) D^t_q(\omega_{12})
\end{eqnarray}
with
\begin{gather}
[\hat D^s_q(\omega_n)]^{-1}=q^2+\frac{16}{\sigma_{xx}}(z+\Gamma_2
- 2 \hat \Gamma) \omega_n\\ D^{-1}_q(\omega_n,\tau_1,\tau_2) =
D^{-1}_q(\omega_n) + i \frac{8 z_v
\Delta_v}{\sigma_{xx}}(\tau_1-\tau_2),\\
[D^{t}_q(\omega_n,\tau_1,\tau_2)]^{-1} = [D^{t}_q(\omega_n)]^{-1}+
i\frac{8 z_v \Delta_v}{\sigma_{xx}} (\tau_1-\tau_2).
\end{gather}
In the same way as in Sec.~\ref{Sec:Short:Rel}, the conservation
of the total valley isospin guarantees the relation $z_v =
z+\Gamma_2$. The conservation of the $z$-component of the total
spin, $\rho^+ - \rho^-$, implies that $z_s = 4 \Gamma_{+-}$ (see
Eq.~\eqref{SS1}). Therefore,
\begin{gather}
\frac{d\ln \Gamma_{+-}}{d\xi} = 0 \label{G+-cons}
\end{gather}
for the length scales $L_s\ll L \ll L_v$.

\subsection{One-loop approximation \label{Sec:Intermed:OneLoop}}


Evaluation of the conductivity according to Eq.~\eqref{SigmaODef} in the one-loop
approximation yields
\begin{gather}
\sigma^\prime_{xx}(i\omega_n) = \sigma_{xx}
+\frac{2^8\pi}{\mathbb{D}\sigma_{xx}} \int_p p^2
T\sum_{\omega_m>0} \min\left \{\frac{\omega_m}{\omega_n},1\right
\} D_p^t(\omega_m)\notag
\\ \times D_p(\omega_m+\omega_n)\Bigl [ \sum_{\sigma=\pm} \left (\hat\Gamma \hat
D^s_p(\omega_m)\right )_{\sigma\sigma} - 4 \Gamma_2
D_p(\omega_m)\Bigr ].
\end{gather}
Hence, we find
\begin{gather}
\sigma^\prime_{xx}(i\omega_n) = \sigma_{xx}
+\frac{2^7\pi}{\mathbb{D}\sigma_{xx}}\int_p p^2 T\sum_{\omega_m>0}
\min\{\frac{\omega_m}{\omega_n},1\} D_p(\omega_m) \notag
\\\times D_p(\omega_m+\omega_n) \Bigl [ z D^s_p(\omega_m) -6 \Gamma_2 D^t_p(\omega_m)\notag \\
\hspace{1cm} - (z+2\Gamma_2-4\Gamma_{++}) \tilde{D}^t_p(\omega_m)
\Bigr ]\label{SR}
\end{gather}
where
\begin{gather}
[\tilde D^{t}_q(n)]^{-1} = q^2 +\frac{64}{\sigma_{xx}}
\omega_n\Gamma_{+-} .
\end{gather}
Performing the analytic continuation to the real frequencies,
$i\omega_n\to\omega+i0^+$ in Eq.~\eqref{SR}, one obtains the DC
conductivity in the one-loop approximation:
\begin{gather}
\sigma^\prime_{xx} = \sigma_{xx} -
\frac{2^8\pi}{\mathbb{D}\sigma_{xx}} \int_p p^2 \int_0^\infty
d\omega \,D^2_p(\omega) \Bigl [ z D^s_p(\omega) \notag
\\ -6 \Gamma_2
D^t_p(\omega) - (z+2\Gamma_2-4\Gamma_{++}) \tilde{D}^t_p(\omega)
\Bigr ]\label{sigma2}
\end{gather}

In order to  compute $z^\prime$ and $z_v^\prime$ we have to
evaluate the thermodynamic potential $\Omega$ in the presence of
the finite valley splitting $\Delta_v$. In the one-loop approximation we find
\begin{gather}
T^2\frac{\partial\Omega/T}{\partial T}=  8 N_r T\sum_{\omega_n>0}
\omega_n
 \Bigl [z+\frac{4}{\sigma_{xx}} \int_p \Bigl [
2\Gamma_{+-}\tilde D_p^t(\omega_n)\notag
\\ -(z+\Gamma_2) D_p^t(\omega_n)
+(z+\Gamma_2) \sum_{\tau_1,\tau_2}D_p^t(\omega_n,\tau_1,\tau_2)
\notag
\\-z
\sum_{\tau_1,\tau_2} D_p(\omega_n,\tau_1,\tau_2)\Bigr ]\Bigr
].\label{z1}
\end{gather}
Following definitions~\eqref{Defz} and \eqref{Defzsv} of the
physical observables, we obtain from Eq.~\eqref{z1}
\begin{gather}
z^\prime =z +\frac{8}{\sigma_{xx}}(2\Gamma_2-\Gamma_{++}) \int_p
D_p(0) \label{z2}
\end{gather}
and
\begin{gather}
z_v^\prime = z_v \Biggl [1+4\pi \left (\frac{16
}{\sigma_{xx}}\right )^3(z+\Gamma_2)T\sum_{\omega_n>0} \omega_n
\int_p\Bigl [z D_p^3(\omega_n)\notag \\ -(z+\Gamma_2)
D_p^{t3}(\omega_n)\Bigr ] \Biggr ].\label{zv2}
\end{gather}

We mention that the results~\eqref{sigma2}, \eqref{z2},
\eqref{zv2} can be obtained with the help of the background field
procedure~\cite{Background} applied to the
action~\eqref{SstartSigma1}-\eqref{SstartF1}.

\begin{figure}[t]
  \includegraphics[width=80mm]{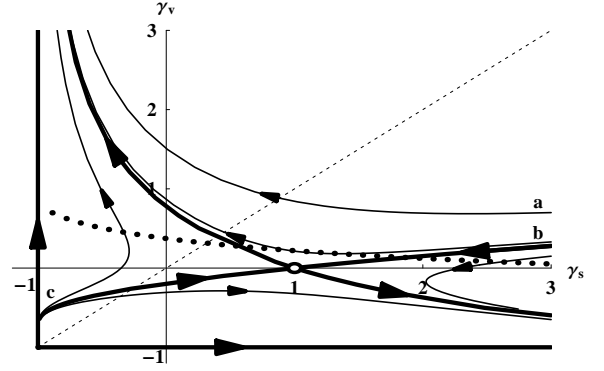}
  \caption{\label{Figure2}
  The projection of the RG flow in the three dimensional
  parameter space $(\sigma_{xx},\gamma_v,\gamma_s)$ onto
  $(\gamma_v,\gamma_s)$ plane for the $SU(2)\times SU(2)$ symmetry case
  (Eqs.~\eqref{RG1_1}-\eqref{RG1_3}). Dots denote the line at which $1+6f(\gamma_v)+f(\gamma_s)=0$.
  The dashed line indicates the line $\gamma_v=\gamma_s$ (see text).
}
\end{figure}

\begin{figure}[t]
  \includegraphics[width=80mm]{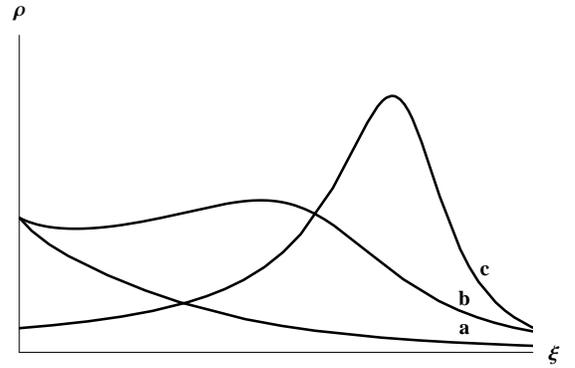}
  \caption{\label{Figure3}
  Schematic dependence of the resistance $\rho=1/(\pi\sigma_{xx})$ on $\xi$. Curves $a$,
  $b$, and $c$ corresponds to the flow lines $a$, $b$, and $c$
  in Fig.~\protect\ref{Figure2} (see text).
  }
\end{figure}

%
%
%
\subsection{One loop RG equations\label{Sec:Intermed:1loopRG}}

Using the standard method,~\cite{Amit} we derive from
Eqs.~\eqref{sigma2}, \eqref{z2} and \eqref{zv2} one-loop results
for the RG equations which determine the $T=0$ behavior of the
physical observables with changing the length scale $L$. It is
convenient to define $\gamma_v = \Gamma_2/z$ and
$\gamma_s=-1+4\Gamma_{+-}/z$. Then, for $\mathbb{D}=2$ we obtain
\begin{gather}
\frac{d \sigma_{xx}}{d\xi} =- \frac{2}{\pi}\left [ 1+6
f(\gamma_{v}) + f(\gamma_{s}) \right ] \label{RG1_1}\\
\frac{d\gamma_v}{d\xi} =
\frac{1+\gamma_v}{\pi\sigma_{xx}}(1+2\gamma_v-\gamma_s)\label{RG1_2}\\
\frac{d\gamma_s}{d\xi} =
\frac{1+\gamma_s}{\pi\sigma_{xx}}(1-6\gamma_v-\gamma_s)\label{RG1_3}\\
\frac{d\ln z}{d\xi} = -\frac{1}{\pi\sigma_{xx}} \left
[1-6\gamma_v-\gamma_s\right ].\label{RG1_4}
\end{gather}
Eqs.~\eqref{RG1_1}-\eqref{RG1_4} constitute one of the main
results of the present paper and describe the system at the
intermediate length scales $L_s\ll L \ll L_v$. We mention that the
length scale $l$ involved in $\xi=\ln L/l$ is now of the order of
$L_s$.

In Figure~\ref{Figure2} we present the projection of the RG flow
in the three dimensional parameter space
$(\sigma_{xx},\gamma_v,\gamma_s)$ onto $(\gamma_v,\gamma_s)$
plane. There is the unstable fixed point at $\gamma_v=0$ and
$\gamma_s=1$. However, for the physical system considered the
fixed point is inaccessible since an initial point of the RG flow
is always situated near the line $\gamma_v=\gamma_s$. As shown in
Fig.~\ref{Figure3}, there are possible three distinct types of the
$\rho(\xi)$ behavior for such initial points. Along the RG flow
line $a$ (Fig.~\ref{Figure2})  that crosses the curve $d$
described by the equation $1+6f(\gamma_v)+f(\gamma_s)=0$ the
resistance demonstrates the metallic behavior: $\rho$ decreases as
$\xi$ grows. If we move along the RG flow line $b$ which
intersects the curve $d$ twice, then the resistance develops the
minimum and the maximum. At last, the resistance on the RG flow
line $c$ which has single crossing with the curve $d$ has the
maximum. Remarkably, in all three cases, the behavior of the
 resistance is of the metallic type for relatively large $L$. The reason of this metallic behavior can be understood
 from the following arguments. At large $\xi$, the coupling $\gamma_v$ flows to large positive values whereas
 $\gamma_s\to -1$. Then, $\Gamma_{+-}/\Gamma_{++} \sim 1/\gamma_v \ll 1$ and the RG Eqs.~\eqref{RG1_1}-\eqref{RG1_4}
 transforms into equations for the single valley system with the conductance equal $\sigma_{xx}/2$.
The metallic behavior of this system  is
well-known.~\cite{Finkelstein}

\section{Completely symmetry broken case\label{Sec:Low}}

\subsection{Effective action\label{Sec:Low:Effect}}

At the long length scales $L\gg L_v$ the symmetry breaking term
$\mathcal{S}_{vb}$ becomes important. In the quadratic approximation
it reads
\begin{equation}
\mathcal{S}_{vb} = \frac{i z_v\Delta_v}{2}\int d^2\mathbf{r}
\sum^{\alpha_j,\sigma}_{n_j,\tau_j}
 \left (\tau_2-\tau_1\right
 )[w_{n_1n_2;\tau_1\tau_2}^{\alpha_1\alpha_2}]_\sigma
 [\bar{w}_{n_2n_1;\tau_2\tau_1}^{\alpha_2\alpha_1}]_\sigma.
\end{equation}
Hence, the modes in $[Q^{\alpha\beta}_{nm;\tau_1\tau_2}]_\sigma$
with $\tau_1\neq \tau_2$ acquire a finite mass of the order of
$z_v\Delta_v$. Therefore, they are negligible at long length
scales $L\gg L_v$. As the result, the matrix $Q$ becomes diagonal
matrix in the valley isospin space. The valley susceptibility
remains constant under the action of the renormalization group on
these length scales:
\begin{equation}
\frac{d z_v}{d \xi} = 0,\qquad L\gg L_v.
\end{equation}
Let us define
\begin{equation}
Q_j^{\alpha\beta}=\{[Q_{11}^{\alpha\beta}]_+,[Q_{-1-1}^{\alpha\beta}]_+,
[Q_{11}^{\alpha\beta}]_-,[Q_{-1-1}^{\alpha\beta}]_-\}.
\end{equation}
Then the action $\mathcal{S}= \mathcal{S}_\sigma+\mathcal{S}_F$
reads
\begin{equation}
\mathcal{S}_\sigma = -\frac{\sigma_{xx}}{32} \sum_{j}\int
d^2\mathbf{r}\tr (\nabla Q_j)^2\label{SstartSigma2}
\end{equation}
and
\begin{gather}
\mathcal{S}_F= \pi T\int d^2\mathbf{r}\sum_{j,k}\sum_{\alpha n}\tr
I_n^\alpha Q_j \hat \Gamma_{jk} \tr I_{-n}^\alpha Q_k \\ + 4\pi T
z \sum_{j}\int d^2\mathbf{r}\tr\eta Q_j,
 \label{SstartF2}
\end{gather}
where
\begin{gather}
\hat\Gamma = \begin{pmatrix}
  \Gamma_{++}-\Gamma_2 & \tilde\Gamma_{++} & \tilde{\Gamma}_{+-} & \Gamma_{+-} \\
\tilde\Gamma_{++} & \Gamma_{++}-\Gamma_2& \Gamma_{+-} &
\tilde\Gamma_{+-}\\
  \tilde\Gamma_{+-} & \Gamma_{+-} &\Gamma_{++}-\Gamma_2 &\tilde\Gamma_{++}\\
\Gamma_{+-}&\tilde\Gamma_{+-}
&\tilde\Gamma_{++}&\Gamma_{++}-\Gamma_2
\end{pmatrix} .\label{Gamma4}
\end{gather}
Initially, at the length scale of the order of $L_v$, the coupling
$\tilde\Gamma_{++}=\Gamma_{++}$ and
$\tilde\Gamma_{+-}=\Gamma_{+-}$. However, more general
structure~\eqref{Gamma4} is consistent with the renormalization
group. It is worthwhile to mention that if the matrix $\hat
\Gamma$ is diagonal then the theory~\eqref{SstartSigma2} and
\eqref{SstartF2} would include four copies of the singlet $U(1)$
theory studied in Refs.~[\onlinecite{BPS,BBP}]. The action
\eqref{SstartF2} corresponds to the following low energy part of
the electron-electron interaction Hamiltonian:
\begin{gather}
\mathcal{H}_\textrm{int} = \frac{1}{2} \int d\mathbf{r}
\sum_{\sigma\sigma^\prime;\tau\tau^\prime}\rho^\sigma_\tau\Bigl [
(\Gamma_s)_{\tau\tau^\prime}^{\sigma\sigma^\prime} + \Gamma_t
(t^a)^{\sigma\sigma}_{\tau\tau}(t^a)^{\sigma^\prime\sigma^\prime}_{\tau^\prime\tau^\prime}
\Bigr ] \rho^\sigma_\tau \notag \\ \rho^\sigma_\tau =
\bar{\psi}^\sigma_\tau\psi^\sigma_\tau \label{HintSBD}
\end{gather}

In order to have the invariance under the global rotations
\begin{gather}
Q_j \to e^{i\hat \chi} Q_j e^{-i\hat \chi},\qquad \hat \chi=
\sum_{\alpha n} \chi^\alpha_n I^\alpha_n,
\end{gather}
the following relation has to be fulfilled
\begin{gather}
z + \Gamma_2 - \Gamma_{++}-\tilde\Gamma_{++} = \Gamma_{+-} +
\tilde{\Gamma}_{+-}.
\end{gather}

\subsection{Perturbative expansions\label{Sec:Low:Pert}}

As above, in order to resolve the constraints $Q_j^2=1$, we shall
use the ``square-root'' parameterization for each $Q_j$: $Q_j =
W_j+\Lambda\sqrt{1-W_j^2}$. Then, the propagators are defined by
the theory~\eqref{SstartSigma2} and \eqref{SstartF2} as
\begin{gather}
\langle [w_{n_1n_2}^{\alpha_1\alpha_2}(q)]_j [w_{n_4
n_3}^{\dag\alpha_4\alpha_3;\tau_4\tau_3}(-q)]_k \rangle =
\frac{32}{\sigma_{xx}} \hat{\mathcal{D}}_{jk}, \\
\hat{\mathcal{D}} =
\delta^{\alpha_1\alpha_3}\delta^{\alpha_2\alpha_4}\delta_{n_{12},n_{34}}
\Bigl [\delta_{n_1,n_3} D_q(\omega_{12}) + \frac{32\pi
T}{\sigma_{xx}}\hat \Gamma \delta^{\alpha_1\alpha_2}\notag
\\
\times  D_q(\omega_{12}) \hat D^c_q(\omega_{12})\Bigr ],
\end{gather}
where
\begin{gather}
[\hat
D^c_q(\omega_n)]^{-1}=q^2 +\frac{16}{\sigma_{xx}} (z-\hat\Gamma)\omega_n.
\end{gather}

The conservation of the $z$-components of the total spin,
$\sum_{\sigma\tau}\sigma\rho_\tau^\sigma$, and the total valley
isospin, $\sum_{\sigma\tau}\tau\rho_\tau^\sigma$,  implies (see
Eq.~\eqref{SS1}) that $z_s=2 \Gamma_{+-}+2\tilde{\Gamma}_{+-}$ and
$z_v = 2 \tilde\Gamma_{++}+2\tilde{\Gamma}_{+-}$.  Since, for
$L\gg L_v$ both $z_s$ and $z_v$ are not renormalized, we obtain
\begin{gather}
\frac{d\tilde{\Gamma}_{+-}}{d\xi}=\frac{d\Gamma_{+-}}{d\xi} =
\frac{d\tilde\Gamma_{++}}{d\xi} =0.\label{Gamma3}
\end{gather}
Since, both $\tilde{\Gamma}_{+-}$ and $\Gamma_{+-}$ coincides at
the length scales $L\sim L_v$ and they are not renormalized we
shall not distinguish $\tilde{\Gamma}_{+-}$ and $\Gamma_{+-}$ from
here onwards. If we introduce $\gamma_s$ and $\gamma_v$ such that
$\Gamma_{+-} =\tilde{\Gamma}_{+-} = z(1+\gamma_s)/4$ and
$\tilde{\Gamma}_{++} = z(1+2\gamma_v-\gamma_s)/4$ then both
$\gamma_{s}$ and $\gamma_v$ coincide with the corresponding
couplings of the previous sections at the length scales $L\sim
L_v$.

\subsection{One-loop approximation\label{Sec:Low:1LoopCalc}}

Evaluating the conductivity with the help of Eq.~\eqref{SigmaODef} in the one-loop approximation, we find
\begin{gather}
\sigma^\prime_{xx}(i\omega_n) = \sigma_{xx}
+\frac{2^8\pi}{\mathbb{D}\sigma_{xx}} \int_p p^2 T
\sum_{\omega_m>0} \min\left\{\frac{\omega_m}{\omega_n},1\right
\}\notag
\\ D_p(\omega_m+\omega_n) D_p(\omega_m)\sum_{j}\left (\hat\Gamma
\hat D^c_p(\omega_m)\right )_{jj}.
\end{gather}
Hence,
\begin{gather}
\sigma^\prime_{xx}(i\omega_n) = \sigma_{xx} +\frac{2^8\pi z
}{\mathbb{D}\sigma_{xx}}  \int_p p^2 T \sum_{\omega_m>0} \min\left
\{\frac{\omega_m}{\omega_n},1\right \}\notag \\
D_p(\omega_m+\omega_n) D_p(\omega_m) \left [ {D}^s_p(\omega_m)-2
\gamma_v\bar{D}^t_p(\omega_m)-\gamma_s\tilde{D}^t_p(\omega_m)\right
]\label{SRR}
\end{gather}
where
\begin{gather}
[\bar D^{t}_q(\omega_n)]^{-1} = q^2
+\frac{32}{\sigma_{xx}}(\Gamma_{++} +\tilde\Gamma_{++})\omega_n.
\end{gather}
Performing the analytic continuation to the real frequencies in
Eq.~\eqref{SRR}, we find
\begin{gather}
\sigma^\prime_{xx} = \sigma_{xx} -\frac{2^8\pi z
}{\mathbb{D}\sigma_{xx}} \int_p p^2 \int_0^\infty d\omega
D^2_p(\omega)\notag \Bigl [ {D}^s_p(\omega)\notag \\\hspace{1cm}-2
\gamma_v\bar{D}^t_p(\omega)-\gamma_s\tilde{D}^t_p(\omega)\Bigr
].\label{sigma3}
\end{gather}

As in the previous Section, in order to compute $z^\prime$ we
evaluate the thermodynamic potential in the one-loop
approximation. The result is
\begin{gather}
T^2\frac{\partial\Omega/T}{\partial T}= 8 T N_r z
\sum_{\omega_n>0} \omega_n
 \Bigl [1+\frac{8}{\sigma_{xx}} \int_p \Bigl [
\frac{(1+\gamma_s)}{2}\tilde D_p^t(\omega_n)\notag \\
+(1+\gamma_v) \bar D_p^t(\omega_n) -2 D_p(\omega_n)\Bigr ]\Bigr
].\label{z4}
\end{gather}
Hence, we obtain
\begin{gather}
z^\prime=  z +\frac{16}{\sigma_{xx}}(\Gamma_2-\Gamma_{++}) \int_p
D_p(0).\label{z3}
\end{gather}

We mention~\cite{Background} that the results~\eqref{Gamma3},
~\eqref{sigma3}, and \eqref{z3} can be obtained with the help of
the background field procedure applied to the
action~\eqref{SstartSigma2}-\eqref{SstartF2}.

\begin{figure}[t]
  \includegraphics[width=80mm]{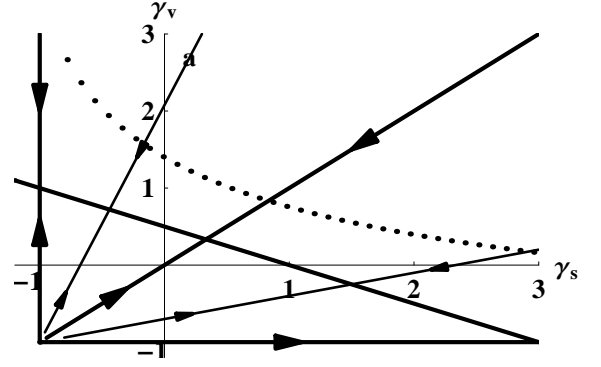}
  \caption{\label{Figure4}
  The projection of the RG flow in the three dimensional
  parameter space $(\sigma_{xx},\gamma_v,\gamma_s)$ onto
  $(\gamma_v,\gamma_s)$ plane for the completely symmetry broken case
  (Eqs.~\eqref{RG3_1}-\eqref{RG3_3}). Dots denote the line at which
  $1+2f(\gamma_v)+f(\gamma_s)=0$ (see text).
}
\end{figure}
\begin{figure}[t]
  \includegraphics[width=80mm]{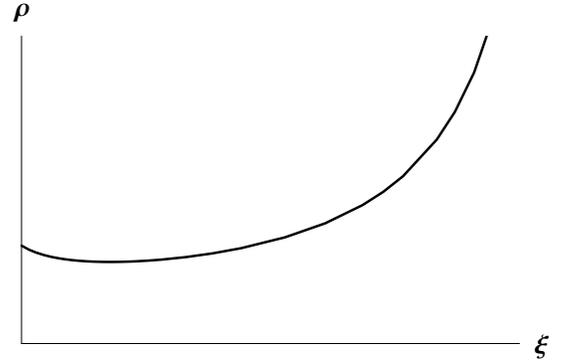}
  \caption{\label{Figure5}
  Schematic dependence of the resistance $\rho=1/(\pi\sigma_{xx})$ on $\xi$ along the
  flow line $a$ in Fig.~\ref{Figure4} (see text).
}
\end{figure}

%
%
%
%
\subsection{One loop RG equations\label{Sec:Low:1loopRG}}

Equations~\eqref{Gamma3}, \eqref{sigma3} and \eqref{z3} allow us to
derive the following one-loop results for the renormalization group functions
which determine the $T=0$ behavior of the physical observables
with changing the length scale $L$ ($\mathbb{D}=2$):
\begin{gather}
\frac{d \sigma_{xx}}{d\xi} =-
\frac{2}{\pi} \left [ 1+2 f(\gamma_{v}) +
f(\gamma_{s}) \right ] \label{RG3_1}\\
\frac{d\gamma_v}{d\xi} =
\frac{1+\gamma_v}{\pi\sigma_{xx}}
(1-2\gamma_v-\gamma_s)\label{RG3_2}\\
\frac{d\gamma_s}{d\xi} =
\frac{1+\gamma_s}{\pi\sigma_{xx}}(1-2\gamma_v-\gamma_s)\label{RG3_3}\\
\frac{d\ln z}{d\xi} = -\frac{1}{\pi\sigma_{xx}} \left
[1-2\gamma_v-\gamma_s\right ]\label{RG3_4}
\end{gather}
The renormalization group equations~\eqref{RG3_1}-\eqref{RG3_4} constitute
one of the main
results of the present paper. We mention that the length scale $l$
involved in $\xi=\ln L/l$ is now of the order of $L_v$ and
Eqs.~\eqref{RG3_1}-\eqref{RG3_4} describe the system at the long
length scales $L \gg L_v$.

The projection of the RG flow for Eqs.~\eqref{RG3_1}-\eqref{RG3_3}
on the $\gamma_v$ -- $\gamma_s$ plane is shown in
Fig.~\ref{Figure4}. There exits the line of the fixed points that
is described by the equation $2\gamma_v+\gamma_s=1$. If the initial
point has large $\gamma_v$ or $\gamma_s$ then the RG flow line
crosses the curve that is determined by the condition
$1+2f(\gamma_v)+f(\gamma_s)=0$. Therefore, the $\rho(\xi)$
dependence along the RG flow line develops the minimum and will be
of the insulating type as is shown in Fig.~\ref{Figure5}.

\section{Discussions and conclusions\label{Sec:Disc}}

The renormalization group equations discussed above describe the
$T=0$ behavior of the observable parameters with changing of the
length scale $L$.  At finite temperatures $T \gg \sigma_{xx}/(z
L^2_{sample})$ where $L_{sample}$ is the sample size, the
temperature behavior of the physical observables can be found from
the RG equations stopped at the inelastic length $L_\textrm{in}$ rather than at the sample size.
Formally, it means that one should substitute $\xi_T= \frac{1}{2}\ln
\sigma_{xx}/(z T l^2)$ for $\xi$ in the RG equations with $\xi_T$ obeying the following
equation~\cite{Euro}
\begin{equation}
\frac{d \xi_T}{d \xi} =1-\frac{1}{2} \frac{d\ln z}{d\xi}.
\label{Lin}
\end{equation}

\begin{figure}[t]
  \includegraphics[width=80mm]{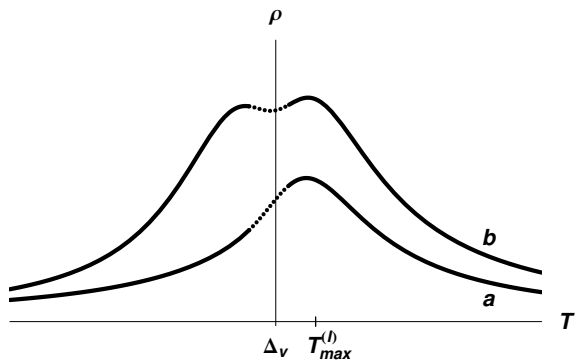}
  \caption{\label{Figure6}
  The schematic $\rho(T)$ dependence in the case of zero parallel magnetic field. See text}
\end{figure}

Having in mind Eq.~\eqref{Lin}, we find that the $T$-behavior of
the resistivity at $B=0$ is described by Eqs.~\eqref{RG0_1} and
\eqref{RG0_2} for $T\gg \Delta_v$ and
Eqs.~\eqref{RG1_1}-\eqref{RG1_3} with interchanged $\gamma_v$ and $\gamma_s$ for $T\ll \Delta_v$.
In what follows, we assume that $\Delta_v < T_\textrm{max}^{(I)}$ where
 $T_\textrm{max}^{(I)}$ denotes the temperature of the maximum point
 that appears in $\rho(T)$ according to the RG Eqs.~\eqref{RG0_1} and
\eqref{RG0_2}. Our assumption is consistent with the experimental
data in Si-MOSFET where, for example,~\cite{PudalovDelta} the
valley splitting is of the order of hundreds of $mK$ and
 $T_\textrm{max}^{(I)}$ is about several Kelvins~\cite{JETPL2}. Then, depending on the
 initial conditions at $T\sim
1/\tau$ two types of the $\rho(T)$ behavior are possible as is
shown in Fig.~\ref{Figure6}. The curve $a$ represents the typical
$\rho(T)$ dependence that was observed in transport experiments on
two-valley 2D electron systems in Si-MOS samples~\cite{Pudalov1}
and n-AlAs quantum well.~\cite{Shayegan3} Surprisingly, the other
behavior with \textit{the two maximum points} is possible, as
illustrated by curve $b$ in Fig.~\ref{Figure6}. So far, this
interesting non-monotonic $\rho(T)$ dependence has been neither
observed experimentally nor predicted theoretically. At very low
temperatures $T\ll \Delta_v$, the metallic behavior of $\rho(T)$
wins even in the presence of the valley splitting.

\begin{figure}[t]
  \includegraphics[width=80mm]{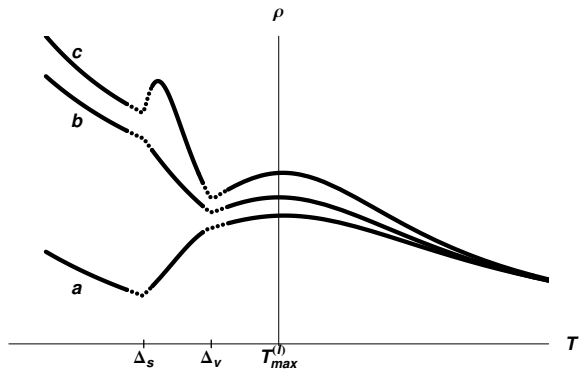}
  \caption{\label{Figure7}
  The schematic $\rho(T)$ dependence in the presence of both spin and valley splitting in the case
  $\Delta_s<\Delta_v$. For the opposite case, the behavior will be
  similar. See text
   }
\end{figure}

In the presence of the sufficiently low parallel magnetic field
$\Delta_s < T_\textrm{max}^{(I)}$, the $\rho(T)$ behavior of
three distinct types is possible as plotted in Fig.~\ref{Figure7}. In all
three cases, the $\rho(T)$ dependence has the maximum point at temperature $T=T_\textrm{max}^{(I)}$ and
is of the insulating type as $T\to 0$.
As follows from
Fig.~\ref{Figure3},
 in the intermediate temperature range, when
$T$ is between $\Delta_s$ and $\Delta_v$, the metallic (curve
$a$), insulating (curve $b$) and nonmonotonic (curve $c$) types of
the $\rho(T)$ behavior emerge. As a result, there has to exist the
$\rho(T)$ dependence with two maximum points in the presence of
$B$.

\begin{figure}[t]
  \includegraphics[width=80mm]{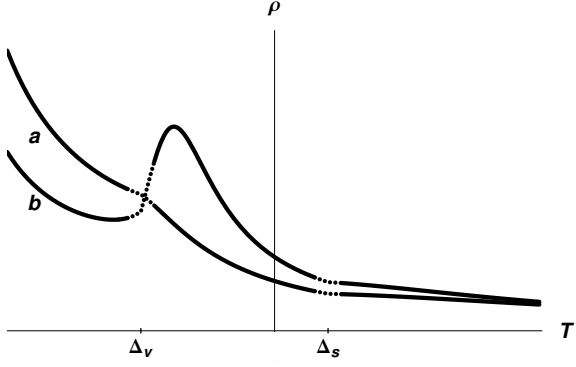}
  \caption{\label{Figure8}
  The schematic $\rho(T)$ dependence in the presence of strong parallel
magnetic field: $\Delta_v , T_\textrm{max}^{(I)}<\Delta_s$.
  See text}
\end{figure}

For high magnetic fields such that $\Delta_s >
T_\textrm{max}^{(I)}$ the maximum point at
$T=T_\textrm{max}^{(I)}$ is absent, and two types of
the $\rho(T)$ behavior are possible as is shown in Fig.~\ref{Figure8}. If
$T_\textrm{max}^{(II)}<\Delta_v$, then the dependence of the resistivity is
monotonic and insulating, see the curve $a$ in
Fig.~\ref{Figure8}. Here, $T_\textrm{max}^{(II)}$ denotes the temperature of the maximum point
 that appears in the resistivity in accord with the RG Eqs.~\eqref{RG1_1} and
\eqref{RG1_2}.
In the opposite case
$T_\textrm{max}^{(II)}>\Delta_v$, a typical $\rho(T)$ dependence is illustrated
by the curve $b$ in Fig.~\ref{Figure8}. Therefore,
if the valley splitting is sufficiently large, i.e., $\Delta_v
> T_\textrm{max}^{(II)}$, then the monotonic insulating behavior
of the resistivity appears in the parallel magnetic field which
corresponds to $\Delta_s \sim T_\textrm{max}^{(I)}$. This is the
case for the experiments on the magnetotransport in
Si-MOSFET.~\cite{Pudalov2,JETPL2} However, if the valley splitting
is small, $\Delta_v < T_\textrm{max}^{(II)}$, then the maximum
point of the $\rho(T)$ dependence survives even in high magnetic
fields but shifts down to lower temperatures.

\begin{figure}[t]
  \includegraphics[width=80mm]{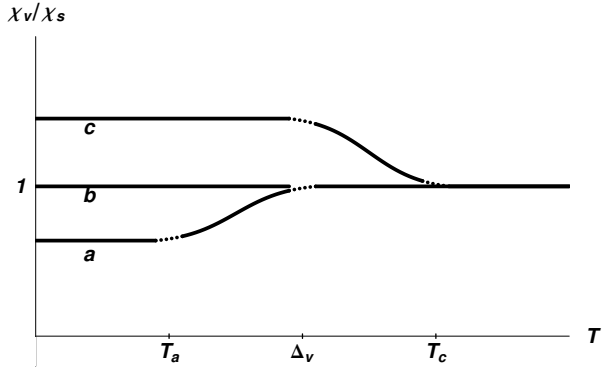}
  \caption{\label{Figure9}
  The schematic dependence of the ration $\chi_v/\chi_s$ on temperature:
  a) for $\Delta_s<\Delta_v$; b) for $\Delta_s=\Delta_v$; c) for
  $\Delta_s>\Delta_v$. The temperature scales $T_{a,c} \equiv
  \Delta_s$.
  }
\end{figure}

In addition, to interesting $T$-dependences of the resistivity,
the theory predicts strong renormalization of the
electron-electron interaction with temperature.
 In order to characterize this renormalization, we consider the ratio
$\chi_v/\chi_s$ of valley and spin susceptibilities. In
Figure~\ref{Figure9}, we present the schematic dependence of
$\chi_v/\chi_s$ on $T$ for a fixed valley splitting but with varying
 spin splitting. At high temperatures, $T\gg \Delta_v, \Delta_s$ the
ratio of the susceptibilities equals unity, $\chi_v/\chi_s=1$. At
low temperatures, $T \ll \Delta_v, \Delta_s$, we find
\begin{equation}
\frac{\chi_v}{\chi_s} =
  \begin{cases}
    < 1 & ,\, \Delta_s < \Delta_v\\
    1 & ,  \,\Delta_s  = \Delta_v \\
    > 1  & ,\,\Delta_s > \Delta_v .
  \end{cases}
\end{equation}
Therefore, the ratio $\chi_v/\chi_s$ at $T\to 0$ is sensitive to
the ratio $\Delta_v/\Delta_s$. This can be used for the
experimental determination of the valley splitting in the 2D
electron system.

Finally, we remind that we do not consider above the contribution
to the one-loop RG equations from the particle-particle (Cooper)
channel. It can be shown (see Appendix) that neither the spin
splitting nor the valley splitting does not change the
``cooperon'' contribution to the RG equations in the one-loop
approximation. Therefore, the Cooper-channel contribution to the
RG equations discussed above can be taken into account by the
substitution of $1+2$ for $1$ in the square brackets of
Eqs.~\eqref{RG0_1}, \eqref{RG1_1} and \eqref{RG3_1}. The
Cooper-channel contribution does not change qualitative behavior
of the resistivity, and the valley and spin susceptibilities
discussed above.

To summarize, we have obtained the novel results on the
temperature behavior of such physical observables as the
resistivity, spin and valley susceptibilities in 2D electron
liquid with two valleys in the MIT vicinity and in the presence of
both the parallel magnetic field and the valley splitting. First,
we found that the metallic behavior of the resistivity at low
temperatures survives in the presence of only the parallel
magnetic field or the valley splitting. If both the spin splitting
and the valley splitting exist then the metallic $\rho$-dependence
crosses over to insulating one at low temperatures. Second, we
have predicted the existence of the novel, nonmonotonic dependence
of resistivity at zero and finite magnetic field in which the
$\rho(T)$ has two maximum points. It would be an experimental
challenge to identify this novel regime.

\begin{acknowledgements}

The authors are grateful to D.A.\,Knyazev, A.A.\,Kuntzevich,
O.E.\,Omelyanovsky, and V.M.\,Pudalov for the detailed discussions
of their experimental data. The research was funded in part by
CRDF, the Russian Ministry of Education and Science, Council for
Grants of the President of Russian Federation, RFBR 07-02-00998-a
and 06-02-16708-a, Dynasty Foundation, Programs of RAS, and
Russian Science Support Foundation.

\end{acknowledgements}

\appendix

\section{``Cooperon" contribution to the conductance}

We start from the standard equation for the ``cooperon''
\begin{gather}
C^{\sigma_1\sigma_2;\tau_1\tau_2}_{\sigma_3\sigma_4;\tau_3\tau_4}(q)
=
\delta^{\sigma_1,\sigma_2}\delta^{\tau_1,\tau_2}
\delta^{\sigma_3,\sigma_4}\delta^{\tau_3,\tau_4}\frac{1}{2\pi\nu\tau_i}
 \notag \\
+I^{\sigma_1\sigma_5;\tau_1\tau_5}_{\sigma_3\sigma_6;\tau_3\tau_6}(q)
C^{\sigma_5\sigma_2;\tau_5\tau_2}_{\sigma_6\sigma_4;\tau_6\tau_4}(q)\label{EqApp1}
\end{gather}
where the ``impurity ladder" is given as
\begin{gather}
I^{\sigma_1,\sigma_2;\tau_1\tau_2}_{\sigma_3\sigma_4;\tau_3\tau_4}(q)
= \frac{1}{2\pi\nu\tau_i} \int_p
G^R_{\sigma_1,\sigma_2;\tau_1\tau_2}(p_+)
G^A_{\sigma_3\sigma_4;\tau_3\tau_4}(p_-)
\end{gather}
Here, $p_\pm=p\pm q/2$ and the impurity averaged Green functions
are given as
\begin{gather}
[G^{R(A)}_{\sigma_1,\sigma_2;\tau_1\tau_2}(p)]^{-1} =
\delta^{\sigma_1,\sigma_2}\delta^{\tau_1\tau_2} \Bigl [
\frac{p^2}{2m_e}-\mu+\Delta_s \sgn\sigma_1\notag \\
\hspace{2cm}+\Delta_v\sgn \tau_1 \pm i/2\tau_i\Bigr ].
\end{gather}
Performing integration, we find for $q\to 0$
\begin{gather}
I^{\sigma_1,\sigma_2;\tau_1\tau_2}_{\sigma_3\sigma_4;\tau_3\tau_4}(q)
=
\delta^{\sigma_1,\sigma_3}\delta^{\sigma_2\sigma_4}
\delta^{\tau_1\tau_3}\delta^{\tau_2\tau_4}  \Bigl [ 1 -D q^2
\tau_i \notag \\ - i \Delta_s \tau_i (\sgn \sigma_3-\sgn\sigma_1)
- i \Delta_v \tau_i (\sgn \tau_3-\sgn\tau_1) \Bigr ].
\end{gather}
Next, solving Eq.~\eqref{EqApp1}, we obtain
\begin{gather}
C^{\sigma_1,\sigma_2;\tau_1\tau_2}_{\sigma_3\sigma_4;\tau_3\tau_4}(q)
=
\delta^{\sigma_1\sigma_2}\delta^{\sigma_3\sigma_4}\delta^{\tau_1\tau_2}\delta^{\tau_3\tau_4}
\frac{1}{2\pi\nu\tau_i^2} \Bigl [ D q^2 \\ +i\Delta_s (\sgn
\sigma_3-\sgn\sigma_1)+i\Delta_v(\sgn\tau_3-\sgn\tau_1)\Bigr
]^{-1} \notag
\end{gather}
The interference correction to the conductance is given as
\begin{gather}
\delta\sigma_{xx} = - \frac{D}{\tau_i} \int_p
G^R(-p)_{\sigma_3\sigma_4;\tau_3\tau_4}
G^A_{\sigma_4\sigma_5;\tau_4\tau_5}(-p) \\ \hspace{1cm}\times
G^R_{\sigma_1\sigma_2;\tau_1\tau_2}(p)
G^A_{\sigma_6\sigma_1;\tau_6\tau_1}(p)  \int \frac{d^2
q}{(2\pi)^2}
C^{\sigma_2\sigma_3;\tau_2\tau_3}_{\sigma_5\sigma_6;\tau_5\tau_6}(q)
. \notag
\end{gather}
Using the result:
\begin{gather}
\int_p G^R(-p)_{\sigma_3\sigma_4;\tau_3\tau_4}
G^A_{\sigma_4\sigma_5;\tau_4\tau_5}(-p)
G^R_{\sigma_1\sigma_2;\tau_1\tau_2}(p)
G^A_{\sigma_6\sigma_1;\tau_6\tau_1}(p)  \notag \\
=
\frac{4\pi\nu\tau_i^3
\delta^{\sigma_3\sigma_5}\delta^{\sigma_2\sigma_6}\delta^{\tau_3\tau_5}\delta^{\tau_2\tau_6}
}{ 1+\Delta^2_s\tau_i^2 (\sgn \sigma_2-\sgn\sigma_3)^2+
\Delta^2_v\tau_i^2 (\sgn \tau_2-\sgn\tau_3)^2},
\end{gather}
we find
\begin{gather}
\delta\sigma_{xx} = - \sum_{\sigma_1\sigma_2}^{\tau_1\tau_2}
\delta^{\sigma_1\sigma_2}\delta^{\tau_1\tau_2} \Bigl [
1+\Delta^2_s\tau_i^2 (\sgn \sigma_1-\sgn\sigma_2)^2\notag \\
+\Delta^2_v\tau_i^2 (\sgn \tau_1-\sgn\tau_2)^2\Bigr ]^{-1}D \int
\frac{d^2 q}{(2\pi)^2} \Bigl [D q^2 \notag \\
+i\Delta_s (\sgn \sigma_1-\sgn\sigma_2)+i\Delta_v (\sgn
\tau_1-\sgn\tau_2)\Bigr ]^{-1} \notag \\
= -\frac{4}{\pi}\int \frac{d q}{q} .
\end{gather}
Therefore, neither the spin splitting nor the valley splitting
affect the Cooper channel (interference) contribution to the
conductance.


\begin{thebibliography}{99}
\bibitem{AFS} T.\,Ando, A.B.\,Fowler, and F.\,Stern,
Rev. Mod. Phys. \textbf{54}, 437 (1982).

\bibitem{Pudalov1} S.V.\,Kravchenko, G.V.\,Kravchenko,
J.E.\,Furneaux, V.M.\,Pudalov, M.\,D'Iorio, Phys. Rev. B
\textbf{50}, 8039 (1994).

\bibitem{prb95}
S.V. Kravchenko,
W.E. Mason, G.E. Bowker, J.E. Furneaux, V.M. Pudalov, and
M.D'Iorio, Phys. Rev. B {\bf 51} 7038 (1995).

\bibitem{Review} E.\,Abrahams, S.V.\,Kravchenko, and
M.P.\,Sarachik, Rev. Mod. Phys. \textbf{73}, 251 (2001);
S.V.\,Kravchenko, and M.P.\,Sarachik, Rep. Prog. Phys,
\textbf{67}, 1 (2004).

\bibitem{Finkelstein} A.M.\,Finkelstein, \textit{Electron liquid in disordered
conductors}, vol. 14 of Soviet Scientific Reviews, ed. by I.M.\,
Khalatnikov, Harwood Academic Publishers, London, (1990).

\bibitem{LargeN} A.\,Punnoose, and A.M.\,Finkelstein, Science
\textbf{310}, 289 (2005).

\bibitem{BPS} M.A.\,Baranov, A.M.M.\,Pruisken, and B.\,\v{S}kori\'{c}, Phys.
Rev. B \textbf{60}, 16821 (1999).

\bibitem{BBP}  M.\,A.\,Baranov, I.\,S.\,Burmistrov, and
A.\,M.\,M.\,Pruisken, Phys. Rev. B \textbf{66}, 075317 (2002).

\bibitem{FP} A.\,Punnoose, and A.M.\,Finkelstein, Phys. Rev. Lett.
\textbf{88}, 016802 (2001).

\bibitem{Pudalov2} D.\,Simonian, S.V.\,Kravchenko,
M.P.\,Sarachik, and V.M.\,Pudalov, Phys. Rev. Lett. \textbf{79},
2304 (1997).

\bibitem{Shayegan0} M. Shayegan, E.P. De Poortere, O. Gunawan, Y.P. Shkolnikov, E. Tutuc, and K. Vakili,
Phys. Stat. Sol.(b) \textbf{243}, 3629 (2006).

\bibitem{Shayegan1} O.\,Gunawan, Y.P.\,Shkolnikov, K.\,Vakili, T. Gokmen, E. P. De
Poortere, and M. Shayegan, Phys. Rev. Lett. \textbf{97}, 186404
(2006).

\bibitem{Shayegan2} O. Gunawan, T. Gokmen, K. Vakili, M. Padmanabhan, E. P. De
Poortere, and M. Shayegan, Nature Phys. \textbf{3}, 388 (2007).

\bibitem{JETPL1} Recently, the $\rho(T)$-behavior in the presence of parallel magnetic field
has been studied in I.S. Burmistrov and N.M. Chtchelkatchev, JETP
Lett. \textbf{84}, 656 (2006). However, due to a mistake we have
found the same RG equations as ones given by Eqs.~\eqref{RG1_1},
\eqref{RG1_3} and \eqref{RG1_4} but with $\gamma_s=\gamma_v$. This
has led us to the erroneous conclusion that in the presence of the
parallel magnetic field only the $\rho(T)$-dependence crosses over
from metallic to insulating.



\bibitem{Iordansky} S.\,Brener, S.V.\,Iordanski, and A.\,Kashuba,
Phys. Rev. B \textbf{67}, 125309 (2003).

\bibitem{Kuntsevich} A.Yu.\,Kuntsevich, N.N.\,Klimov, S.A.\,Tarasenko,
N.S.\,Averkiev, V.M.\,Pudalov, H.\,Kojima, M.E.\,Gershenson, Phys.
Rev. B \textbf{75}, 195330 (2007).

\bibitem{KirkpatricBelitz} D.\,Belitz and T.R.\,Kirkpatrick, Rev. Mod. Phys. \textbf{66}, 261 (1994).

\bibitem{AleinerZalaNarozhny} G.\,Zala, B.N.\,Narozhny, and I.L.\,Aleiner, Phys. Rev. B \textbf{64}, 214204 (2001).

\bibitem{Castellani} C.\,Castellani, C.\,Di Castro, P.A.\,Lee, and
M.\,Ma, Phys. Rev. B \textbf{30}, 527 (1984).

\bibitem{FreeElectrons} F.\,Wegner, Z. Phys. B \textbf{35}, 207
(1979); L.\,Schaefer and F.\,Wegner, Z. Phys. B \textbf{38}, 113
(1980); A.J.\,McKane and M.\,Stone, Ann. Phys. (N.Y.)
\textbf{131}, 36 (1981); K.B.\,Efetov, A.I.\,Larkin,
D.E.\,Khemel'nitzkii, Sov. Phys. JETP \textbf{52}, 568 (1980).

\bibitem{CasDiCas} C.\,Castellani and C.\,Di Castro, Phys. Rev. B
\textbf{34}, 5935 (1986).

\bibitem{Aronov-Altshuler} B.L. Altshuler and A.G. Aronov, in
{\it Electron-Electron Interactions in Disordered Conductors}, ed.
A.J. Efros and M. Pollack, Elsevier Science Publishers,
North-Holland, 1985.

\bibitem{Unify} A.M.M.\,Pruisken, M.A.\,Baranov,
and B.\,\v{S}kori\'{c}, Phys. Rev. B \textbf{60}, 16807 (1999);

\bibitem{KamenevAndreev} A. Kamenev and A. Andreev, Phys. Rev. B \textbf{60}, 2218
(1999).

\bibitem{CastelChi} C.\,Castellani, C.\,Di Castro, P.A.\,Lee, M.\,Ma,
S.\,Sorella, and E.\,Tabet, Phys. Rev. B \textbf{33}, 6169 (1986).

\bibitem{RecentProgress} A.M.M.\,Pruisken and I.S.\,Burmistrov,
Ann. of Phys. (N.Y.) {\bf 322}, 1265 (2007).


\bibitem{Meerovich} A.E.\,Meyerovich, JLTP \textbf{53}, 487 (1983).

\bibitem{Zala} G.\,Zala, B.N.\,Narozhny, I.L.\,Aleiner, and V.I.\,Falko, Phys. Rev. B \textbf{69}, 075306 (2004).

\bibitem{Background} I.S.\,Burmistrov and N.M.\,Chtchelkatchev, unpublished.

\bibitem{Amit} D.J.\,Amit, \textit{Field theory, renormalization
group, and critical phenomena}, (World Scientific, 1984).




\bibitem{Euro} A.M.M.\,Pruisken and M.A.\,Baranov, EuroPhys.
Lett. \textbf{31}, 543 (1995).

\bibitem{PudalovDelta} N.\,Klimov, M.E.\,Gershenson, H.\,Kojima,
D.A.\,Knyazev, V.M.\,Pudalov, to be published.

\bibitem{JETPL2} D.A.\,Knyazev,
O.E.\,Omel'yanovskii, V.M.\,Pudalov, and I.S.\,Burmistrov, JETP
Lett. \textbf{84}, 662 (2006).

\bibitem{Shayegan3} S.J.\,Papadakis, and M.\,Shayegan, Phys. Rev. B \textbf{57}, R15068 (1998).

\bibitem{Cohen} M.H.\,Cohen and A.M.M.\,Pruisken,
Phys. Rev. B {\bf 49}, 4593 (1994).



\end{thebibliography}
\end{document}